\documentclass[pra,aps,twocolumn,showpacs]{revtex4-1}

\usepackage{amsmath}
\usepackage[pdftex]{graphicx}
\usepackage{epstopdf}

\usepackage{bm}

\newcommand{\kk}{\bm{q}}

\begin{document}

\author{Krzysztof Gawarecki}
\author{Pawe{\l} Machnikowski}
\affiliation{Institute of Physics, Wroc{\l}aw University of
Technology, 50-370 Wroc{\l}aw, Poland}

\title{Electron states, phonon-assisted relaxation and tunneling in self-assembled
  quantum dot molecules in an electric field}

\begin{abstract}
We present a theoretical analysis of the phonon-assisted relaxation in
a system composed of two self-assembled vertically stacked quantum
dots. We construct realistic model, which takes into account the
geometry and strain distribution in the system. We calculate
phonon-assisted relaxation rates between the two lowest states (in
one- and two-electron cases). The relaxation rates and energy levels
are studied as a function of external (axial) electric field and
geometry of the structure (dot sizes). We show that the relaxation
times can be as low as 1~ps but
efficent relaxation occurs only for very finely tuned dots.
\end{abstract}


\maketitle

\section{Introduction}

Coupled quantum dots are not only interesting from the theoretical
point of view but also promising for designing quantum-coherent
devices, including spin-based quantum bits. However, 
phonon-related processes are inevitable in a cristal environment and
may limit the feasibility of implementing quantum control in these
systems. On the other hand, in many applications (e.g., lasers) fast
relaxation between two states is desired. Therefore, it is essential
to understand not only the spectral properties of various kinds of
semiconductor nanostructures but also phonon-induced processes.

Carrier transfer and relaxation in self-assembled structures was observed in many experiments. In those works, optical spectroscopy methods were used 
in stacked QDMs
\cite{heitz98,reischle07,nishibayashi08,park07,bajracharya07,sales04,%
mazur05,seufert01,tackeuchi00,ortner05c,chang08,gerardot05,nakaoka06}. Miscellaneous mechanisms could be responsible for the observed properties. The heterogeneity of considered structures and posible transfer mechanisms is reflected in a relatively wide distribution of  the measured transfer rates. However, in most cases the kinetics is attributed to
tunneling  \cite{reischle07,nakaoka04,mazur05,seufert01,heitz98,tackeuchi00,%
rodt03,chang08,nakaoka06}. 

In recent works (Refs. \cite{pochwala08,gawarecki10}), we studied phonon-induced relaxation in one- and two-particle systems. In Ref. \cite{gawarecki10}, we investigated the dependence of the relaxation on the system geometry.
In this contribution, we also take into account dependence on external (axial) electric field for one- and two-particle systems.
We show that phonon-induced relaxationmay be controlled either by
system geometry or by external electric field. 

\section{Model}
\label{sec:model}

The system under consideration is composed of two vertically stacked QDs, formed in the Stransky--Krastanov self-assembly process. 
The dots are modeled as spherical domes of hights $H_{1}$ and $H_{2}$
(the index ``1'' refers to the lower dot) and base radii $r_{1}$ and
$r_{2}$. The domes rest on wetting layer of thickness
$H_{\mathrm{WL}}$ and are separated by a distance D
(base-to-base). The details of the model and the parameter values are
given in Ref.~ \cite{gawarecki10}. 
We consider InAs dots in a GaAs enviroment.  Approximately, our system
has an axial symetry, so it is treated as cylindrically
symmetric. This assumption strongly simplifies our calculations.   
The lattice constants mismatch leads to
the appearance of strain. The strain distribution is represented by
displacement fields that are found by minimizing the elastic energy of
the system \cite{pochwala08}.   

States of an electron in the strained
nanostructure in the present approach are determined by the position-dependent 
conduction band edge
and by the effective masses, which also vary across the structure.
The local band structure is obtained from the 8-band
Hamiltonian with strain-induced terms (Bir-Pikus
Hamiltonian) using the L\"owdin elimination
\cite{pochwala08,gawarecki10}. From this, we calculate 
the conduction band edge and the effective mass tensor as a function of
the spatial position. The electron wave functions are calculated within a
variational multi-component envelope function approximation for
equation with the Hamiltonian \cite{pochwala08}
\begin{eqnarray*}
H & = & -\frac{\partial}{\partial x}
\frac{\hbar^{2}}{2m_{\bot}(\rho,z)} \frac{\partial}{\partial x}
-\frac{\partial}{\partial y}
\frac{\hbar^{2}}{2m_{\bot}(\rho,z)} \frac{\partial}{\partial y} \\
&&-\frac{\partial}{\partial z}
\frac{\hbar^{2}}{2m_{z}(\rho,z)} \frac{\partial}{\partial z}
+E_{\mathrm{c}}(\rho,z)-e E_{el} z,
\end{eqnarray*}
where $m_{\bot}(\rho,z)$,$m_{z}(\rho,z)$ are  the components of the
effective mass tensor, $E_{\mathrm{el}}$ is electrical field
magnitude, 
$E_{\mathrm{c}}(\rho,z)$ is the conduction band edge and $e$ is electron charge. 

Next, we derive two-particle states within
the standard configuration-interaction approach in a restricted basis 
of lowest-energy configurations (only two lowest single-electron
states are included). The Hamiltonian of the interacting two-electron
system has the form \cite{gawarecki10} 
\begin{equation}\label{ham-coulomb}
H =  \sum_{n,\sigma} \epsilon_{n} a_{n\sigma}^{\dag}a_{n\sigma}
+\frac{1}{2} \sum_{ijkl} \sum_{\sigma,\sigma'} 
v_{ijkl}
a_{i\sigma}^{\dag}a_{j\sigma'}^{\dag}  a_{k\sigma'} a_{l\sigma},
\end{equation}
where 
\begin{eqnarray}
v_{ijkl} & = & 
\frac{e^{2}}{4 \pi \varepsilon_{0} \varepsilon_{\mathrm{r}}} \int d^{3}\bm{r}
\int d^{3}\bm{r}' \nonumber \\
&&\times \psi_{i}^{*}(\bm{r}) \psi_{j}^{*}(\bm{r}') 
\frac{1}{| \bm{r}-\bm{r}'|}
\psi_{k}(\bm{r}') \psi_{l}(\bm{r}). \label{v-ijkl}
\end{eqnarray}
Here $\psi_{i}^{*}(\bm{r})$ are single-particle eigenfunctions, $\varepsilon_{0}$ is the vacuum
permittivity, and $\varepsilon_{\mathrm{r}}$ is the dielectric constant of
the semiconductor. After diagonalizing the Hamiltonian, we obtain three lowest singlet states.

Finally, we derive the phonon-assisted relaxation between the
single-electron states as well as between the two lowest two-electron
states. The coupling between the electrons and phonons is described by the
Hamiltonian
\begin{equation}\label{ham-e-ph}
H_{\mathrm{e-ph}}=\sum_{nm,\sigma} a_{n,\sigma}^{\dag}
a_{m,\sigma}^{\phantom{\dag}}
\sum_{s,\kk} F_{s,nm}(\kk) 
\left(b_{s,\kk}^{\phantom{\dag}} + b_{s,-\kk}^{\dag} \right),
\end{equation}
where $F_{s,nn'}(\kk)$ is the coupling constants.  
We use the Fermi golden rule to calculate the relaxation rates between
the lowest energy eigenstates. 
In the calculation of phonon-induced relaxation rates, we take into
account carrier-phonon couplings via deformation 
potential and piezoelectric interactions. 

\section{Results}

\begin{figure}[tb]
\begin{center}
\includegraphics[width=85mm]{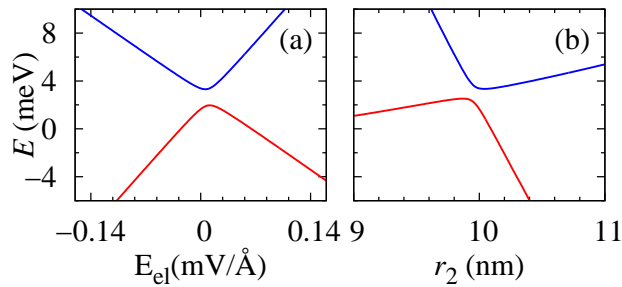}
\end{center}
\caption{\label{fig:one-electron}(a) The single-electron
  energy levels  with a fixed size of both of dots (base radius $r_1 =
  r_{2} = 10$~nm) as a function of the electric field.  
(b) The single-electron
  energy levels  with a fixed size of the
  lower dot as a function of the size (base radius $r_{2}$) of the
  upper one. Here $r_{1}=10$~nm, $H_{1}=3.7$~nm,
  $H_{2}/r_{2}=0.37$. The energy reference level is 0.8~eV above the
  conduction band edge of unstrained bulk InAs. In both cases, the
  dot separation is $D=12$~nm.} 
\end{figure}

The single particle energy levels obtained within our approach for the
fixed distance $D=120$~nm between the dots are shown in
Fig.~\ref{fig:one-electron}(a-b).   
In Fig.~\ref{fig:one-electron}a the energy is shown as a function of
the external electrical field. Here, both dots have the same size.  
On the other hand, in Fig.~\ref{fig:one-electron}b, we show energy
dependence on the upper dot size. The size of the lower dot is kept
constant, while the base radius $r_{2}$ of the upper dot and its
height $H_{2}$ are varied, with $H_{2}/r_{2}$ constant. 
Electronic (tunnel) coupling between the dots leads to the appearance
of an anticrossing pattern. In Fig.~\ref{fig:one-electron}b it occurs
near the point where the dots become equal. In
Fig.~\ref{fig:one-electron}a the resonance is located near the point
$E_{\mathrm{el}}=0$. 
\begin{figure}[tb]
\begin{center}
\includegraphics[width=85mm]{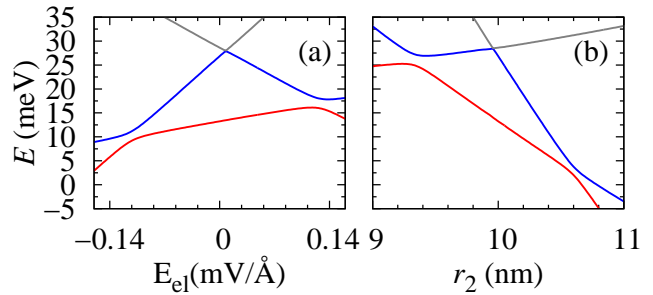}
\end{center}
\caption{\label{fig:two-electron} The two-electron
  energy levels as a function of the electrical field (a) and of the upper dot size (b). The system parameters used in the calculations are identical as in the one-electron case. The energy reference level is 1.6~eV above the
  conduction band edge of unstrained bulk InAs.}
\end{figure}

The spectrum of a two-electron system  is dominated by antricrossings
of different occupation configurations \cite{szafran01,szafran05}. 
In Fig.~\ref{fig:two-electron} the three lowest spin-singlet
eigenstates of the interacting 
two-electron system are shown. In a similar way as in the one electron
case, we present the dependence on the external electrical field
(Fig.~\ref{fig:two-electron}b) and on the upper dot size
(Fig.~\ref{fig:two-electron}a). 
The central resonance occurs when the doubly occupied configurations 
have similar energy. This anticrossing is very narrow (less than
$0.1$~meV). The other two anticrossings occur at the
degeneracy point between the singly occupied 
configuration and configurations with two electrons in
the larger dot. These two splittings are wider
than those appearing between the single-electron states, shown in
Fig.~\ref{fig:one-electron} (2~meV vs. 1.5~meV). It is because the anticrossing of
two-electron configurations is enhanced by Coulomb terms
\cite{grodecka08a}. 
\begin{figure}[tb]
\begin{center}
\includegraphics[width=85mm]{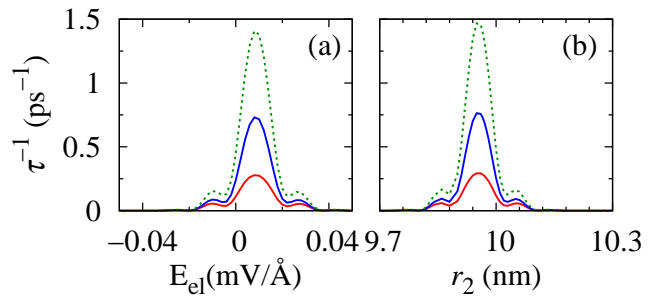}
\end{center}
\caption{\label{fig:tun-one-electron} 
  The thermalization rate as a function of the electrical field (a) and of the upper dot size (b). Results for a one electron system in a structure with
  $r_{1}=10$~nm, $H_{1}=3.7$~nm and $H_{2}/r_{2}=0.37$ at $T=0$~K (red solid
  line), 20~K (blue dashed line), and 40~K (green dotted line).} 
\end{figure}

The single particle relaxation rates are shown in
Fig.~\ref{fig:tun-one-electron} as functions of 
the upper dot size (Fig.~\ref{fig:tun-one-electron}b) and the
electrical field (Fig.~\ref{fig:tun-one-electron}a). These results are
calculated for the same sample 
geometries as in Fig.~\ref{fig:one-electron} at three different
temperatures. The overall magnitude of the
relaxation rates depends on the spatial overlap between the wave
functions corresponding to the states involved in the
transition. This is why the rates are large at the resonance and become
smaller when the system is shifted from the resonance point. 
\begin{figure}[thb]
\begin{center}
\includegraphics[width=85mm]{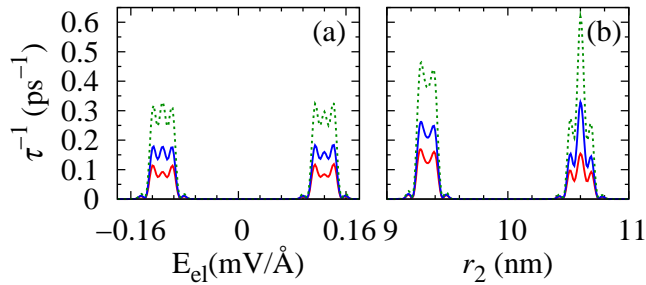}
\end{center}
\caption{\label{fig:tun-two-electron}
  The thermalization rate between the two lowest two-electron states at $T=0$~K (red solid
  line), 20~K (blue dashed line), and 40~K (green dotted line). Plots show the thermalization as a function of the electrical field (a) and of the upper dot size (b).} 
\end{figure}
Figs.~\ref{fig:tun-two-electron}a,b show the relaxation rates for a two electron system.
We use the same sample geometries as in the one-electron case and investigate the dependence on electrical field and upper dot size.
In general, the relaxation rates are similar to those found in the single
electron case. This is due to the fact that both these processes are physically very similar. The only difference is that  
in the single electron case it tunnels towards an empty QD, whereas
in the two electron case, there is already another electron. 
It leads to shifts (due to Coulomb interaction) to the
parameter regimes where  energies compensates for the Coulomb
repulsion.  

\section{Conclusion}
We have studied the phonon-assisted relaxation (thermalization) for
single-electron and two-electron configurations in self-assembled
quantum dots. Our model has taken into account the geometry and strain
distribution in a QDM. 
The results show that the phonon relaxation is very efficient
in a very narrow range of parameters near the anticrossings of energy
levels. We conclude that relaxation times on the order of at least hundreds of picoseconds should be typical. 
The results for an electric field and various upper dot sizes are quantitatively consistent.  
Therefore, it is possible to control phonon--induced relaxation by system geometry as well as by an external electric field.

\begin{acknowledgments}
This work was supported by the TEAM programme of the Foundation for
Polish Science co-financed from the European Regional Development
Fund.
\end{acknowledgments}

\bibliographystyle{prsty}

\end{document}